\documentclass[11pt,a4paper]{article}
\usepackage[hyperref]{acl2018}
\usepackage{times}
\usepackage{latexsym}
\usepackage{amsmath}
\usepackage{graphicx}
\usepackage{tabu}
\usepackage{float}
\usepackage{booktabs}
\usepackage{algorithm}
\usepackage[noend]{algpseudocode}
\usepackage{multirow}
\usepackage{enumitem}
\usepackage{multicol}
\usepackage[linewidth=1pt]{mdframed}
\usepackage{booktabs}
\usepackage{graphicx}
\usepackage{url}
\usepackage{tikz}
\usepackage{lipsum}

\aclfinalcopy 
\def\aclpaperid{57} 

\newcommand\Tstrut{\rule{0pt}{2.3ex}}         
\newcommand\Bstrut{\rule[-0.9ex]{0pt}{0pt}}
\title{Alignment Analysis of Sequential Segmentation of Lexicons to Improve Automatic Cognate Detection}

\author{Pranav A \\
  Big Data Institute \\
  Hong Kong University of Science and Technology \\
  Hong Kong \\
  {\tt cs.pranav.a@gmail.com} }

\date{}

\begin{document}
\maketitle
\begin{abstract}
    Ranking functions in information retrieval are often used in search engines to recommend the relevant answers to the query. 
    This paper makes use of this notion of information retrieval and applies onto the problem domain of cognate detection. 
    The main contributions of this paper are: (1) positional segmentation, which incorporates the sequential notion; (2) graphical error modelling, which deduces the transformations.
    The current research work focuses on classification problem; which is distinguishing whether a pair of words are cognates.
    This paper focuses on a harder problem, whether we could predict a possible cognate from the given input.
    Our study shows that when language modelling smoothing methods are applied as the retrieval functions and used in conjunction with positional segmentation and error modelling gives better results than competing baselines, in both classification and prediction of cognates.
\end{abstract}

\section{Introduction}

Cognates are a collection of words in different languages deriving from the same origin. 
The study of cognates plays a crucial role in applying comparative approaches for historical linguistics, in particular, solving language relatedness and tracking the interaction and evolvement of multiple languages over time. 
A cognate instance in Indo-European languages is given as the word group: \textit{night} (English), \textit{nuit} (French), \textit{noche} (Spanish) and \textit{nacht} (German).

The existing studies on cognate detection involve experiments which distinguish between a pair of words whether they are cognates or non-cognates \cite{ciobanu2014building, List:2012:LAD:2388655.2388671}.
These studies do not approach the problem of predicting the possible cognate of the target language, if the cognate of the source language is given.
For example, given the word \textit{nuit}, could the algorithm predict the appropriate German cognate within the huge German wordlist?
This paper tackles this problem by incorporating heuristics of the probabilistic ranking functions from information retrieval.
Information retrieval addresses the problem of scoring a document with a given query, which is used in every search engine. 
One can view the above problem as the construction of a suitable search engine, through which we want to find the cognate counterpart of a word (query) in a lexicon of another language (documents). 

This paper deals with the intersection between the areas of information retrieval and approximate string similarity (like the cognate detection problem), which is largely under-explored in the literature. 
Retrieval methods also provide a variety of alternative heuristics which can be chosen for the desired application areas \cite{Fang:2011:DEI:1961209.1961210}.
Taking such advantage of the flexibility of these models, the combination of approximate string similarity operations with an information retrieval system could be beneficial in many cases.
We demonstrate how the notion of information retrieval can be incorporated into the approximate string similarity problem by breaking a word into smaller units.
Regarding this, Nguyen et al. \shortcite{ngubook} has argued that segmented words are a more practical way to query large databases of sequences, in comparison with conventional query methods. 
This further encourages the heuristic attempt at imposing an information retrieval model on the cognate detection problem in this paper.

Our main contribution is to design an information retrieval based scoring function (see section 4) which can capture the complex morphological shifts between the cognates.
We tackled this by proposing a shingling (chunking) scheme which incorporates positional information (see section 2) and a graph-based error modelling scheme to understand the transformations (see section 3).
Our test harness focuses not only on distinguishing between a pair of cognates, but also the ability to predict the cognate for a target language (see section 5).

\section{Positional Character-based Shingling}
This section examines on converting a string into a shingle-set which includes the encodings of the positional information.
In this paper, we notify, $S$ as the shingle-set of cognate from the source language and $T$ as the shingle-set of cognate for the target language. 
The similarity between these split-sets is denoted by $S \cap T$.
An example of cognate from the source language, $S$ (Romanian) could be shingle set of the word \textit{rosmarin} and $T$ (Italian) could be \textit{romarin}.

\textbf{K-gram shingling:} 
Usually, set based string similarity measures are based on comparing overlap between the shingles of two strings.
Shingling is a way of viewing a string as a document by considering $k$ characters at a time.
For example, the shingle of the word \textit{rosmarin} is created with $k = 2$ as: $S = \left\lbrace \langle \textit{s} \rangle \textit{r, ro, os, sm, ma, ar, ri, in, n} \langle / \textit{s} \rangle \right\rbrace$. 
Here, $\langle \textit{s} \rangle$ is the start sentinel token and $\langle / \textit{s} \rangle$ is the stop sentinel token.
For the sake of simplicity, we have ignored sentinel tokens; which transforms into: $S = \left\lbrace \textit{r, ro, os, sm, ma, ar, ri, in, n} \right\rbrace$. 
This method splits the strings into smaller $k$-grams without any positional information.

\subsection{Positional Shingling from 1 End}
We argue that the unordered \textit{k-grams} splitting could lead to an inefficient matching of strings since a shingle set is visualized as the bag-of-words method. 
Given this, we propose a positional k-gram shingling technique, which introduces position number in the splits to incorporate the notion of the sequence of the tokens. 
For example, the word \textit{rosmarin} could be position-wise split with $k = 2$ as: $S = \left\lbrace \textit{1r, 2ro, 3os, 4sm, 5ma, 6ar, 7ri, 8in, 9n} \right\rbrace$. 

Thus, the member \textit{4sm} means that it is the fourth member of the set.
The motivation behind this modification is that it retains the positional information which is useful in probabilistic retrieval ranking functions.

\subsection{Positional Shingling from 2 Ends}
The main disadvantage of the positional shingling from single end is that any mismatch can completely disturb the order of the rest, leading to low similarity.
For example, if the query is \textit{rosmarin} with cognate \textit{romarin}, the corresponding split sets would be $\left\lbrace \textit{1r, 2ro, 3os, 4sm, 5ma, 6ar, 7ri, 8in, 9n} \right\rbrace$ and $\left\lbrace \textit{1r, 2ro, 3om, 4ma, 5ar, 6ri, 7in, 8n} \right\rbrace$.
The order of the members after \textit{2ro} is misplaced, thus this will lead to low similarity between two cognates.
Only $\left\lbrace \textit{1r, 2ro} \right\rbrace$ is common between the cognates.
Considering this, we propose positional shingling from both ends, which is robust against such displacements.

\begin{figure}[h]
	\centering
	\includegraphics[width=0.4\textwidth]{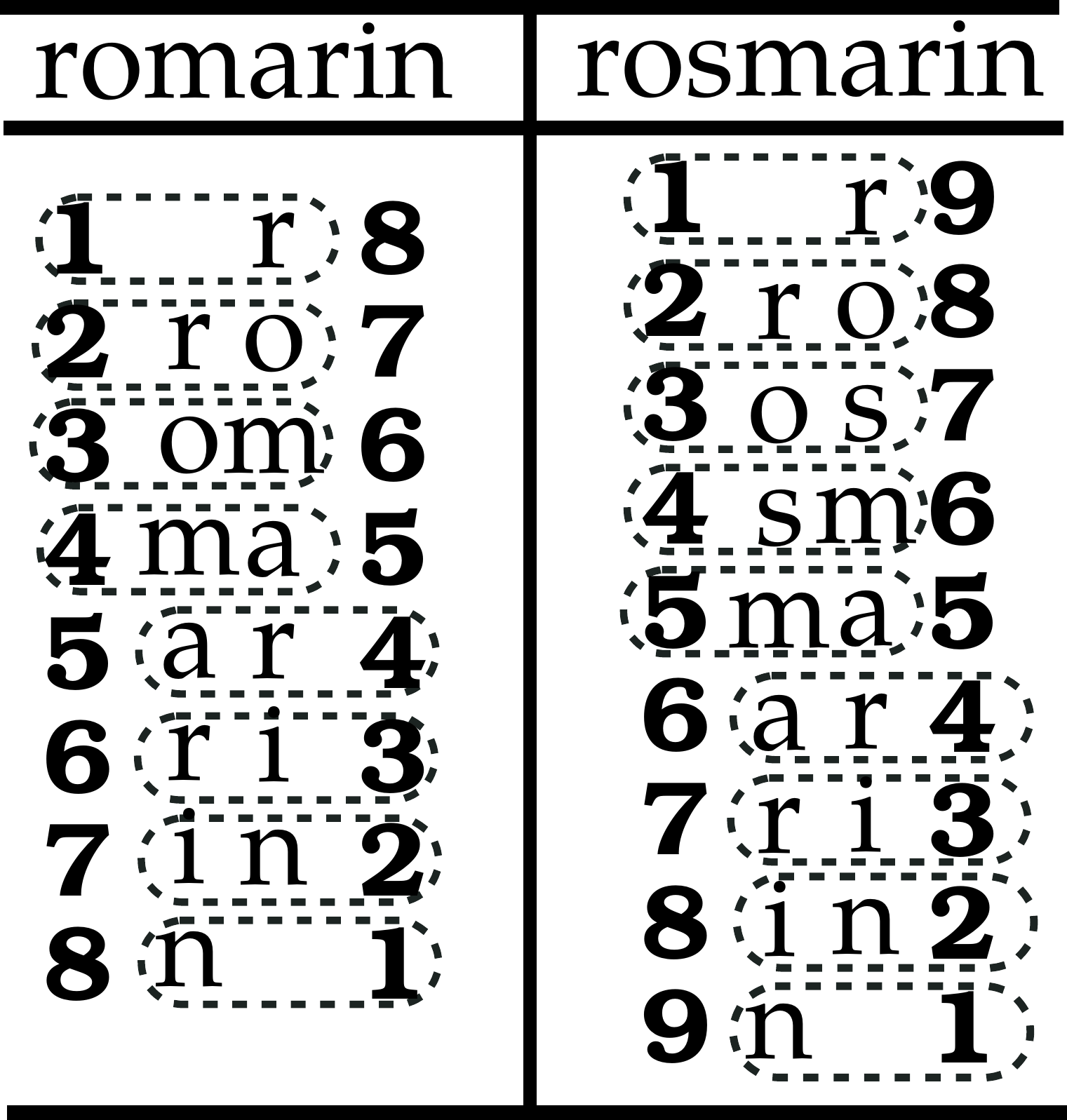}
	\caption{The process of positional tokenisation from both ends. On the left, algorithm segments the Romanian word \textit{romarin} into the split-set $\left\lbrace \textit{1r, 2ro, 3om, 4ma, ar4, ri3, in2, n1} \right\rbrace$. On the right, the algorithm segments \textit{rosmarin} into $\left\lbrace \textit{1r, 2ro, 3os, 4sm, 5ma, ar4, ri3, in2, n1} \right\rbrace$. }
	\label{algo}
\end{figure}

We attach a position number to the left if the numbering begins from the start, and to the right if the numbering begins from the end.
Then the smallest position number is selected between the two position numbers.
If the position numbers are equal, then we select the left position number as a convention.
Figure \ref{algo} gives an exemplification of this algorithm illustrated with splits of \textit{romarin} and \textit{rosmarin}.

In Figure 1, split sets of \textit{rosmarin} and \textit{romarin} are shown. After taking intersection of them, we get $\left\lbrace \textit{1r, 2ro, ar4, ri3, in2, n1} \right\rbrace$, indicating a higher similarity.

\section{Graphical Error Modelling}


Once shingle sets are created, common overlap set measures like set intersection, Jaccard \cite{jarvelin2007s}, XDice \cite{XDice} or TF-IDF \cite{tfidf} could be used to measure similarities between two sets.
However, these methods only focus on similarity of the two strings.
For cognate detection, it is crucial to understand how substrings are transformed from source language to target language.
This section discusses on how to view this "dissimilarity" by creating a graphical error model. 

%
%

Algorithm 1 explicates the process of graphical error modelling.
	For illustration purposes, we visualize the procedure via a Romanian-Italian cognate pair (\textit{mesia}, \textit{messia}). 
	If the source language is Romanian, then $S = \left\lbrace \textit{1m, 2me, 3es, si3, ia2, a1} \right\rbrace$, which is the split-set of \textit{mesia}.
	Let the target language by Italian. Then the split-set of the Italian word \textit{messia}, denoted as $T$, will be $\left\lbrace \textit{1m, 2me, 3es, 4ss, si3, ia2, a1} \right\rbrace$.
	Thus $|S \cap T|$ is the number of common terms.
	Thus the term matches are, 
	$S \cap T = \left\lbrace \textit{1m, 2me, 3es, si3, ia2, a1} \right\rbrace$.
	We are interested in examining the "dissimilarity", which are the leftover terms in the sets.
	That means, we need to infer a certain pattern from leftover sets, which are $S - \lbrace S \cap T \rbrace$ and $T - \lbrace S \cap T \rbrace$.
	Thus we can draw mappings to gather information of the corrections.
	Let \textit{top} and \textit{bottom} be the \textbf{ordered sets} referring to $S - \lbrace S \cap T \rbrace$ and $T - \lbrace S \cap T \rbrace$ respectively.
	Referring to the example, $T - S \cap T = \left\lbrace \textit{4ss} \right\rbrace$, a \textit{bottom} set.
	Similarly, $S - S \cap T = \left\lbrace \right\rbrace$, a \textit{top} set.
	Then we follow instructions given in algorithm 1.

\Tstrut	
	\hrule

\noindent \Tstrut\textbf{Algorithm 1}: Graphical Error Model\Bstrut

\hrule
\Tstrut
\noindent \textbf{Graphical Error Model} takes two split sets generated by the shingling variants, namely \textit{top} and \textit{bottom}.
The objective is to output a graphical structure showing connections between members of the \textit{top} and the \textit{bottom} sets.

\begin{figure}[h]
	\centering
	\includegraphics[width=0.4\textwidth]{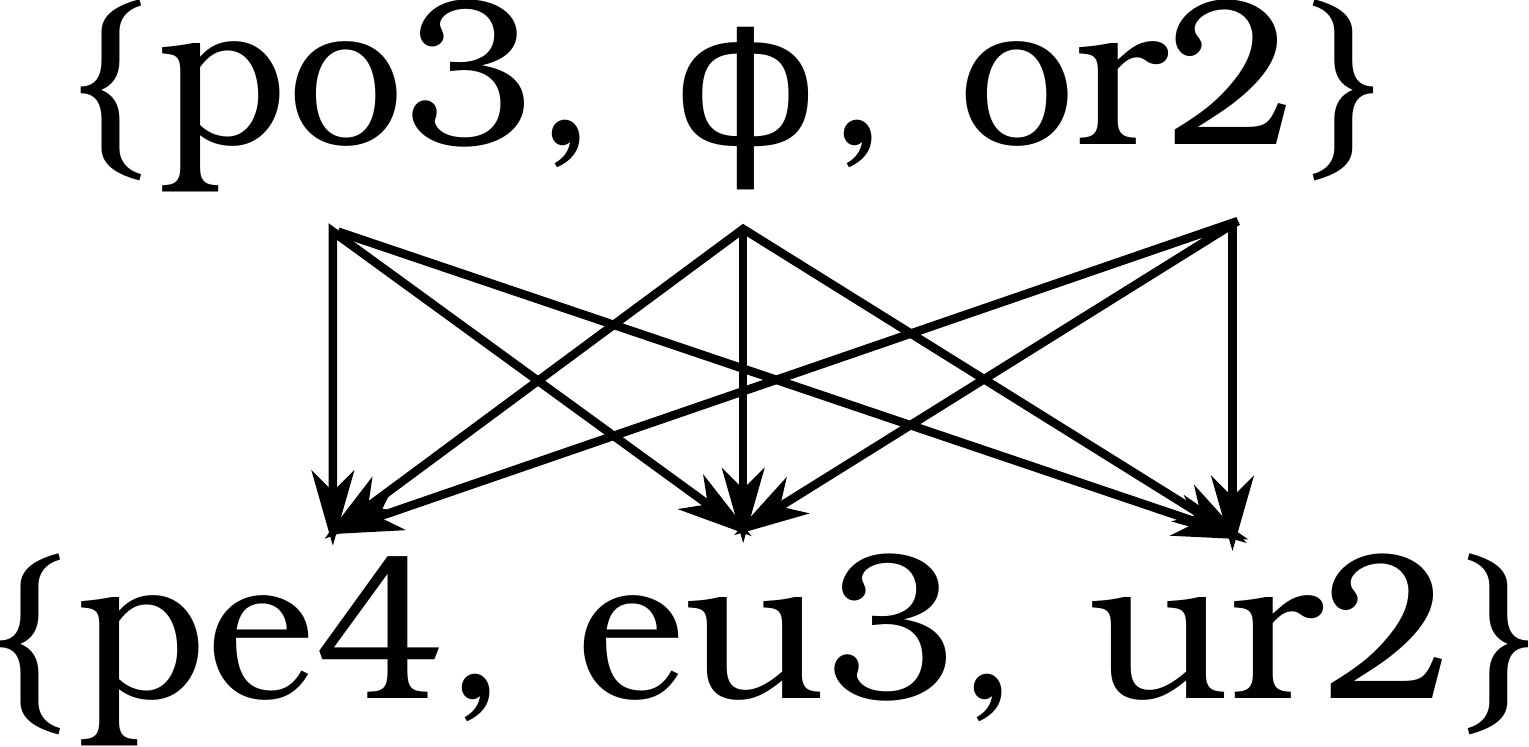}
	\caption{The figure shows the bipartite graph output of the algorithm when the source cognate is \textit{stupor} and the target cognate is \textit{stupeur}.}
	\label{graph}
\end{figure}

\begin{enumerate}[nolistsep]
	\item \textbf{Initialization of the \textit{top} and \textit{bottom}:}
	If the given sets \textit{top} and \textit{bottom} are empty, we initialize them by inserting an empty token ($\phi$) into those sets.
	

	\textbf{Running example:} This step transforms \textit{top} set as $\left\lbrace \phi \right\rbrace$ and \textit{bottom} as $\left\lbrace \textit{4ss} \right\rbrace$.
	
	\item \textbf{Equalization of the set cardinalities:}
	The cardinalities of the sets \textit{top} and \textit{bottom} made equal by inserting empty tokens ($\phi$) into the middle of the sets.
	
	\textbf{Running example:} The set cardinalities of \textit{top} and \textit{bottom} were already equal.
	Thus the output of this step are \textit{top} set as $\left\lbrace \phi \right\rbrace$ and \textit{bottom} as $\left\lbrace \textit{4ss} \right\rbrace$. 
	
	\item \textbf{Inserting the mappings of the set members into the graph:}
	The empty graph is initialized as $graph = \lbrace \rbrace$.
	The directed edges are generated, originating from every set member of \textit{top} to every set member of \textit{bottom}.
	This results in a complete directed bipartite graph between \textit{top} and \textit{bottom} sets.
	Each edge is assigned a probability $P(e)$ which is discussed in a later section.
	
	\textbf{Running example:} The output of this step would be complete directed bipartite graph between \textit{top} and \textit{bottom} sets which is $\left\lbrace \phi \rightarrow \textit{4ss} \right\rbrace$
	
	One more example is provided in figure \ref{graph}.
	
	\Tstrut
%
%
%
%
%
%
%
%
\end{enumerate}\hrule\Tstrut 

	\textbf{Intuition: }
	The edges created as the result of this graph could be used for probabilistic calculations which are detailed more in section \ref{error}.
	Intuitively, $\phi \rightarrow \textit{4ss}$ means that if the letter \textit{s} is added at position 4 of the word of the source \textit{mesia}, then one could get the target word \textit{messia}.

    \section{Evaluation Function}
    
    The design of our evaluation function focuses on two main properties: set based similarity (see section \ref{sim}) and probabilistic calculation through graphical model (see section \ref{error})
    
    \subsection{Similarity Function}
    \label{sim}
    Usually, the computation of similarity between two sets is done by metrics like Jaccard, Dice and XDice \cite{XDice}.
    Dynamic programming based methods like edit distance and LCSR (Longest Common Subsequence Ratio, \cite{lcsr}) are also often used to calculate similarity between two strings.
    Ranking functions incorporate more complex but necessary features which are needed to distinguish between the documents.
    
In this paper, we use BM25 and a Dirichlet smoothing based ranking function to compute the similarity.
BM25 considers term-frequency, inverse document frequency and length normalization based penalization features for similarity calculations.
Dirichlet smoothing function \cite{Robertson:2009:PRF:1704809.1704810} makes use of language modelling features and tunable parameter which aids in Bayesian smoothing of unseen shingles in the split sets \cite{Blei:2003:LDA:944919.944937}.

 \subsection{Error Modelling Function}
 \label{error}
 
 The information of the common morphological transformations for cognates between two different languages helps in determining if a pair of words could be cognates.
    Based on the graphs of cognate pairs between Italian and Romanian (section 3), which models the morphological shifts between the cognates in the two languages, we define an error modelling function on any pair of words from the two languages. 
    The split set from the source language is denoted by $S$ and target language by $T$, then probabilistic function would be:
    \begin{equation*}
	\begin{aligned}
	\pi \textit{(S, T)} = \frac{1}{|G(S,T)|} \sum_{e \in G(S,T)} \textit{P(e)}^q
	\end{aligned}
	\end{equation*}
	where $G(S,T)$ is the constructed graph of $S$ and $T$, the strength parameter is called $q$ here with the range of $(0,\infty)$, and $P(e)$ is the probability of edge $e$ to occur in between two cognates, which is estimated by its frequency of being observed in the graphs of cognate pairs in the training set.
	
	\begin{figure}[h]
		\centering
		\includegraphics[width=0.3\textwidth]{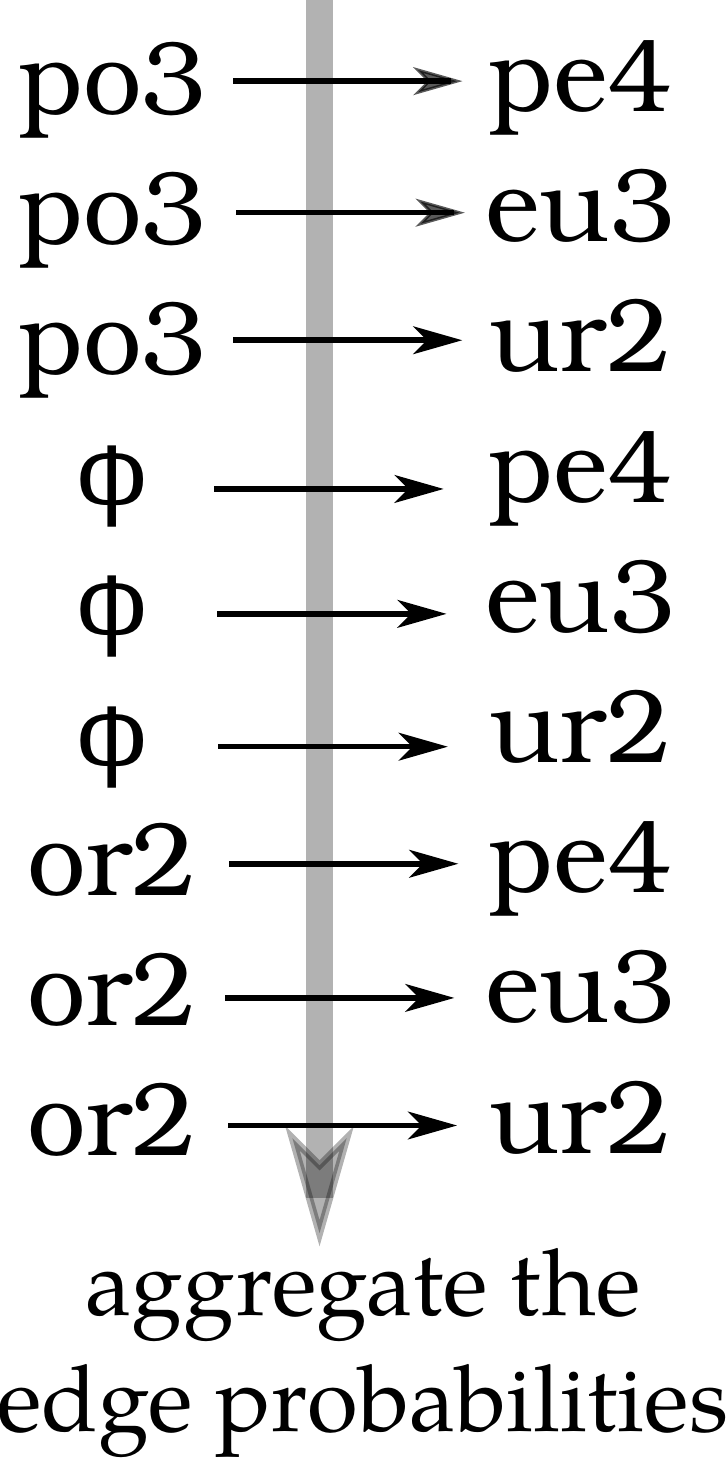}
		\caption{From the graph created in figure \ref{graph}, we calculate the probabilities of each edge (by computing frequencies and smoothing) and then aggregate all the probabilities of edges in the graph.}
		\label{adding}
	\end{figure} 
	
\begin{figure}[!htp]
		\centering
		\includegraphics[width=0.3\textwidth]{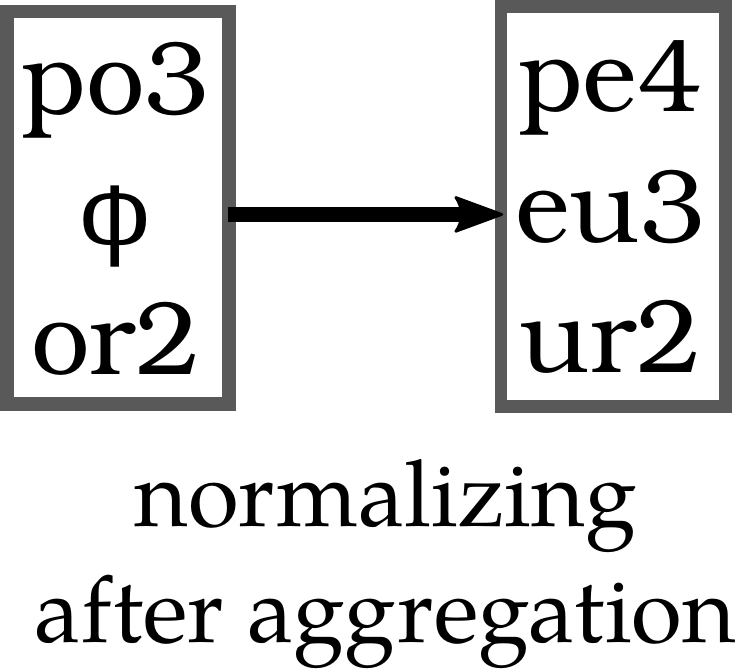}
		\caption{After aggregating, we normalize the sum and the graph conversion score is outputted.}
		\label{final}
	\end{figure}

	Figure \ref{adding} illustrates the aggregation of edges in the graph and figure \ref{final} shows the final output of the error modelling function after normalizing.
	
	$\pi \textit{(S, T)}$ is called the error modelling function defined for the word pair, which is an intuitive calculation of probabilty between a pair of cognates through estimating their transformations.
	$q$ is a tunable parameter that controls the effect of the probabilistic frequencies $P(e)$ observed in the training set, often useful in avoiding overfitting. 
	$\frac{1}{|G(S,T)|}$ is the normalization factor to allow us to compare the quantity across different word pairs.
	
	
\subsection{Combining Error Modelling and Similarity Function metrics}
\label{balance}
In this subsection, we merge the notion of similarity and dissimilarity together.
 We combine a set-based similarity function (discussed in section \ref{sim}) and the error modelling function (discussed in section \ref{error}) into a score function by a weighted sum of them, which is,
	\begin{equation*}
	\label{main}
	\textit{score}(S, T) =  \lambda \times \textit{sim(S, T)} + (1 - \lambda) \times \pi \textit{(S, T)}
	\end{equation*}
	where $\lambda \in [0, 1]$ is a weight based hyperparameter, $sim(S, T)$ is a set-based similarity between $S$ and $T$, and $\pi (S, T)$ is the graphical error modelling function defined above.


\section{Test Harness}

Table \ref{my-label} summarizes the results of the experimental setup.
The elements of test harness are mentioned as following:

\begin{table*}[htp]
\centering
\resizebox{\textwidth}{!}{%
\begin{tabular}{@{}lllllllllll@{}}
\toprule
\multicolumn{3}{c}{\multirow{2}{*}{\textbf{Algorithms}}} & \multicolumn{2}{c}{\textbf{Ro - It}} & \multicolumn{2}{c}{\textbf{Ro - Fr}} & \multicolumn{2}{c}{\textbf{Ro - Es}} & \multicolumn{2}{c}{\textbf{Ro - Pt}} \\ \cmidrule(l){4-11} 
\multicolumn{3}{c}{} & \multicolumn{1}{c}{Acc} & \multicolumn{1}{c}{MRR} & \multicolumn{1}{c}{Acc} & \multicolumn{1}{c}{MRR} & \multicolumn{1}{c}{Acc} & \multicolumn{1}{c}{MRR} & \multicolumn{1}{c}{Acc} & \multicolumn{1}{c}{MRR} \\ \midrule
\multicolumn{3}{c}{Edit Distance \cite{lcsr}} & 0.53 & 0.11 & 0.52 & 0.13 & 0.58 & 0.15 & 0.54 & 0.13 \\
\multicolumn{3}{c}{XDice \cite{XDice}} & 0.54 & 0.19 & 0.53 & 0.14 & 0.59 & 0.16 & 0.53 & 0.14 \\
\multicolumn{3}{c}{SVM with Orthographic Aligment \cite{ciobanu2014building}} & 0.81 & 0.18 & 0.87 & 0.15 & 0.86 & 0.16 & 0.73 & 0.14 \\
\multicolumn{3}{c}{Phonetic Encodings in CNN \cite{cnn}} & 0.69 & 0.21 & 0.78 & 0.17 & 0.77 & 0.19 & 0.66 & 0.15 \\
\multicolumn{3}{c}{Hidden alignment CRF \cite{mccallum2012conditional}} & 0.84 & 0.51 & 0.85 & 0.48 & 0.84 & 0.50 & 0.71 & 0.45 \\ \midrule
\textbf{Shingling Technique} & \textbf{Ranking Function} &  &  &  &  &  &  &  &  \\ \midrule
Bigram, 0-ended & TF-IDF & & 0.54 & 0.18 & 0.52 & 0.14 & 0.59 & 0.15 & 0.55 & 0.11 \\
Bigram, 1-ended & TF-IDF & & 0.59 & 0.19 & 0.54 & 0.18 & 0.60 & 0.18 & 0.57 & 0.12 \\
Bigram, 2-ended & TF-IDF & & 0.64 & 0.25 & 0.63 & 0.21 & 0.68 & 0.22 & 0.65 & 0.17 \\
(Bi + Tri)-gram, 2-ended & TF-IDF & & 0.64 & 0.25 & 0.64 & 0.21 & 0.57 & 0.22 & 0.65 & 0.18 \\
Bigram, 2-ended  & BM25 & & 0.84 & 0.40 & 0.87 & 0.37 & 0.86 & 0.34 & 0.73 & 0.35 \\
Bigram, 2-ended & Dirichlet & & 0.84 & 0.41 & 0.86 & 0.38 & 0.86 & 0.39 & 0.74 & 0.36 \\
Bigram, 2-ended & BM25 + Graphical Error Model & & 0.87 & 0.64 & 0.89 & 0.51 & 0.86 & 0.54 & 0.78 & 0.55 \\
Bigram, 2-ended & Dirichlet + Graphical Error Model & & \textbf{0.88} & \textbf{0.67} & \textbf{0.89} & \textbf{0.59} & \textbf{0.87} & \textbf{0.60} & \textbf{0.80} & \textbf{0.58} \\ \bottomrule
\end{tabular}%
}
\caption{Results on the test dataset. The upper half denotes the baselines used and the lower half describes our ablation experiments. For the experiment 1, we evaluate using the accuracy (Acc) metric. We used MRR (Mean Reciprocal Rank) for describing the second experiment. Higher scores signify the better performance. The maximum value possible is 1.0. It is worth noting that for the classification problem (experiment 1), our algorithm has slight improvement. However for the recommendation problem (experiment 2), our algorithm shows massive improvement. The thresholds, hyper-parameters, source code and sample Python notebooks are available at our github repository: \url{https://github.com/pranav-ust/cognates}}
\label{my-label}
\end{table*}

\subsection{Setup Description}

\textbf{Dataset:}
The experiments in this paper are performed on the dataset used by Ciobanu \textit{et al} \shortcite{ciobanu2014building}.
The dataset consists 400 pairs of cognates and non-cognates for Romanian-French (Ro - Fr), Romanian-Italian (Ro - It), Romanian-Spanish (Ro - Es) and Romanian-Portuguese (Ro - Pt).
The dataset is divided into a 3:1 ratio for training and testing purposes.
Using cross-validation, hyperparameters and thresholds for all the algorithms and baselines were tuned accordingly in a fair manner.

\noindent \textbf{Experiments:}
Two experiments are included in test harness.
\begin{enumerate}[nolistsep]
	\item We provide a pair of words and the algorithms would aim to detect whether they are cognates.
	Accuracy on the test set is used as a metric for evaluation.
	\item We provide a source cognate as the input and the algorithm would return a ranked list as the output.
	The efficiency of the algorithm would depend on the rank of the desired target cognate.
	This is measured by MRR (Mean Reciprocal Rank), which is defined as, $MRR = \sum_{i=1}^{|dataset|} \frac{1}{rank_i}$, where $rank_i$ is the rank of the true cognate in the ranked list returned to the $i^{th}$ query.
	This dataset is prepared by listing search candidates as the entire lexicon of the particular language.
	The target list is the whole lexicon for that particular language. 
	Given the input cognate, the algorithm will output possible matches after evaluating the whole lexicon list.
	Thus, the collection of documents are lexicons (search space) and queries would be the cognates.
\end{enumerate}

\subsection{Baselines}

\textbf{String Similarity Baselines:}
It is intuitive to compare our methods with the prevalent string similarity baselines since the notions behind cognate detection and string similarity are almost similar.
Edit Distance is often used as the baseline in the cognate detection papers \cite{lcsr}. 
This computes the number of operations required to transform from source to target cognate.
We have also incorporated XDice \cite{XDice}, which is a set based similarity measure that operates between shingle set between two strings.
Hidden alignment conditional random fields (CRF) are often used in transliteration which serves as the generative sequential model to compute the probabilities between the cognates, which is analogous to learnable edit distance \cite{mccallum2012conditional}.
Among these baselines, CRF performs the best in accuracy and MRR.

\noindent \textbf{Orthographic Cognate Detection:}
Papers related to this notion usually take alignment of substrings which in classifier like support vector machines \cite{ciobanu2015automatic, ciobanu2014building} or hidden markov models \cite{bhargava2009multiple}.
We included Alina \textit{et al} as the baseline \shortcite{ciobanu2014building}, which employs the dynamic programming based methods for sequence alignment following which features were extracted from the mismatches in the word alignments.
These features are plugged into the classifier like SVM which results in decent performance on accuracy with an average of 84\%, but only 16\% on MRR.
This result is due to the fact that a large number of features leads to overfitting and scoring function is not able to distinguish the appropriate cognate.

\noindent \textbf{Phonetic Cognate Detection:}
Research in automatic cognate identification using phonetic aspects involve computation of similarity by decomposing phonetically transcribed words \cite{kondrak2000new}, acoustic models \cite{mielke2012phonetically}, phonetic encodings \cite{rama2015automatic}, aligned segments of transcribed phonemes \cite{list2012lexstat}.
We implemented Rama's research \shortcite{rama2015automatic}, which employs a Siamese convolutional neural network to learn the phonetic features jointly with language relatedness for cognate identification, which was achieved through phoneme encodings. 
Although it performs well on accuracy, it shows poor results with MRR, possibly the reason as same as SVM performance.

\subsection{Ablation experiments}

We experiment with the variables like length of substrings, ranking functions, shingling techniques, and graphical error model, which are detailed in the Table \ref{my-label}.
Amongst the shingling techniques, we found that character bigrams with 2-ended positioning give better results.
Adding trigrams to the database does not give major effect on the results.
The results clearly indicate that adding graphical error model features greatly improve the test results.
Amongst the ranking functions, Dirichlet smoothing tends to give better results, possibly due to the fact that it requires fewer parameters to tune and is able to capture the sequential data (like substrings) better than other ranking functions \cite{Fang:2011:DEI:1961209.1961210}.
The hyperparameter $\lambda$ mentioned in the section \ref{balance}, was tuned around 0.6, which shows the 60\% contribution by the similarity function and 40\% contribution by the dissimilarity.
Overall, the combination of bigrams with 2-ended positional shingling, graphical error modelling with Dirichlet ranking function gives the best performance with an average of 86\% on accuracy metric and 60\% on MRR.

\section{Conclusions}

We approach towards the harder problem where the algorithm aims to find a target cognate when a source cognate is given.
Positional shingling outperformed non-positional shingling based methods, which demonstrates that inclusion of positional information of substrings is rather important.
Addition of graphical error model boosted the test results which shows that it is crucial to add dissimilarity information in order to capture the transformations of the substrings.
Methods whose scoring functions rely only on complex machine learning algorithms like CNN or SVM tend to give worse results when searching for a cognate, due to huge output space.

\section*{Acknowledgements}

This work would not be possible without the support from my parents.
I would like to thank the NLP community for providing me open-sourced resources to help an underprivileged and naive student like me.
Finally, I would like to thank the reviewers, mentors, and organizers for ACL-SRW for supporting student research.
Special thanks to my classmate Chun Sik Chan and SRW mentor Sam Bowman for providing excellent critiques for this paper, and Alina Ciobanu for providing the dataset.

\newpage

\bibliography{acl2018}
\bibliographystyle{acl_natbib}

\end{document}